\documentclass[11pt,a4paper]{article}

\usepackage{a4wide}
\usepackage{graphicx}
\usepackage{hyperref}

\begin{document}

\setlength{\parindent}{0pt}
\addtolength{\parskip}{6pt}
\newlength{\figWidth} \setlength{\figWidth}{0.5\textwidth}

\author{Kieran Smallbone \\ [24pt]
\emph{Manchester Centre for Integrative Systems Biology} \\
\emph{131 Princess Street, Manchester M1 7DN, UK} \\
\href{mailto:kieran.smallbone@manchester.ac.uk}{\tt kieran.smallbone@manchester.ac.uk}
}

\title{One lump or two?\thanks{This preprint first appeared on Nature Precedings on 1 December 2009 [\href{http://dx/doi.org/10.1038/npre.2009.4033.1}{doi:10.1038/npre.2009.4033.1}].}}

\date{}

\maketitle

\begin{abstract}
We investigate methods for modelling metabolism within populations of cells. Typically one represents the interaction of a cloned population of cells with their environment as though it were one large cell. The question is as to whether any dynamics are lost by this assumption, and as to whether it might be more appropriate to instead model each cell individually. We show that it is sufficient to model at an intermediate level of granularity, representing the population as two interacting lumps of tissue.
\end{abstract}

\section{Introduction}

The emerging field of systems biology seeks to reconcile subcellular-level components (such as enzymatic reactions) with cellular- and organism-level behaviour (such as metabolism). Non-linear processes dominate these interactions; experience from other areas of science has taught us that mathematical models, continuously revised by new information, must be used to describe and interpret complex biological phenomena~\cite{lazebnik02,wiechert02}.

\begin{figure}
\centering
(a) \resizebox{\figWidth}{!}{\includegraphics{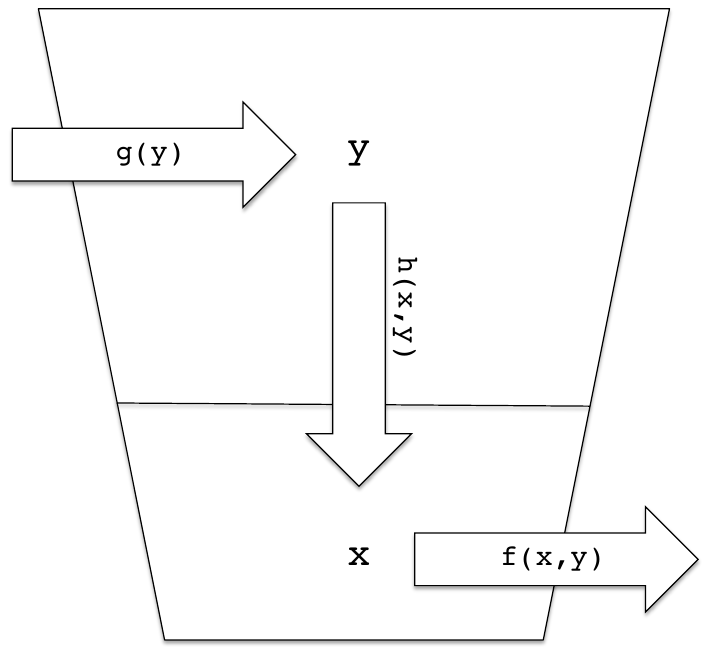}}\\
(b) \resizebox{\figWidth}{!}{\includegraphics{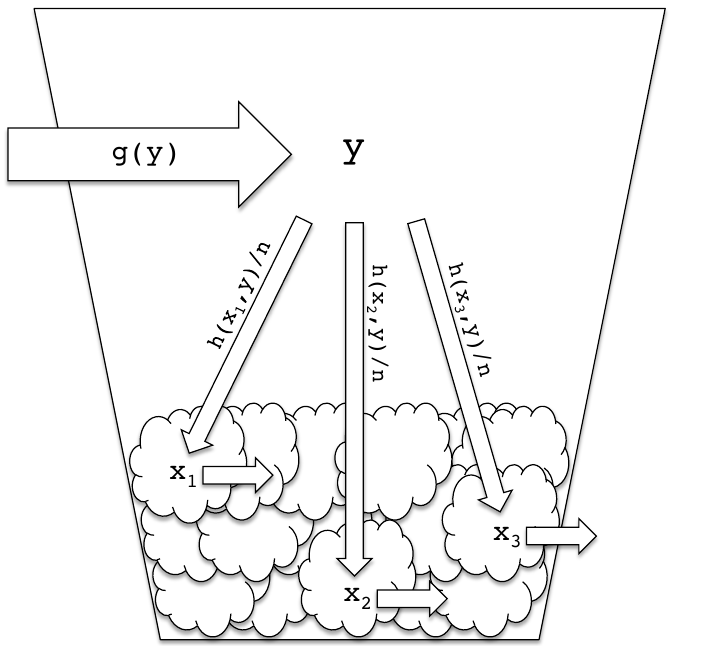}}
\caption{Modelling at different scales. See Eqs.~(\ref{eq1})--(\ref{eq2}) for a mathematical representation, where $y$ denotes extracellular and $x$ intracellular concentrations. Typically, (a) one models the cell population as one bulked compartment; at the other end of the granularity scale, (b) one could consider each of the $n$ cells individually, which would lead to approximately $n$ times as many ODEs.}
  \label{fig0}
\end{figure}

As systems biology grows, so we see a proliferation of mathematical models of cell metabolism and signalling -- see the many examples at the model repositories \href{http://www.ebi.ac.uk/biomodels-main/}{BioModels.net}~\cite{lenovere06} and \href{http://models.cellml.org}{CellML.org}~\cite{lloyd04}. Given the inherent difficulties in in performing single cell experiments, one property held in common by many of these models is the assumption of ``lumped dynamics''. To explain this term, consider a typical scenario in which a million {\em S.~cerevisiae} are grown in a chemostat. Experiments are performed to measure average metabolite concentrations over the population of yeast cells. A mathematical model of metabolism is then built in which Òthe cellÓ has these average characteristics, but a volume equivalent to a million cells (see Fig.~\ref{fig0}).

Given the identical metabolic characteristics of each clonal cell, it would seem natural to approximate the system by lumping the population as a single mass. Intuition would suggest that dynamics are unchanged but, as we shall see below, this linear, verbal reasoning approach is incorrect. However, we show that it is not necessary to consider each individual cell -- which would lead to a million times as many ODEs -- rather correct dynamics can be captured by considering two interacting lumps of cells.

\section{A theorem}

We frame the problem mathematically. Let $x_i$ denote a set of metabolite concentrations within cell $i$, and $y$ a set of external concentrations (see Fig.~\ref{fig0}). Assuming each cell has identical characteristics, we may write

\begin{eqnarray}
x_i^\prime & = & f(x_i,y) \qquad \qquad i=1,\ldots,n \label{eq1}\\
y^\prime & = & g(y) - \frac{1}{n} \sum_i h(x_i,y) \label{eq2}
\end{eqnarray}

Here $f$ denotes intracellular reactions, $h$ transport into cells and $g$ the rate of metabolite supply.

Linearise about a steady-state $x_i = x^\ast$, $y=y^\ast$ to give stability matrix

\begin{equation}
A_n = \left(
\begin{array}{c c c c | c}
f_x & 0 & \cdots & 0 & f_y \\
0 & \ddots & & & \vdots \\
\vdots & & \ddots & & \vdots \\
0 & & & f_x & f_y \\
\hline
-\frac{1}{n} h_x & \cdots & \cdots & -\frac{1}{n} h_x & g_y - h_y
\end{array}
\right)
\end{equation}

We propose that

\begin{equation}
\lambda(A_1) \subseteq \lambda(A_2) = \lambda(A_3) = \ldots
\end{equation}

where $\lambda$ denotes the spectrum. That is, the system bulked into two compartments has the same eigenvalues as the system with three compartments, but more than the system with one compartment.

To show $\lambda(A_n) \subseteq \lambda(A_{n+1})$, let $v_n = (x_1,\ldots,x_n | y)'$ and suppose $A_n v_n = \lambda v_n$. Taking

\begin{equation}
v_{n+1} = \left( \left. x_1, \ldots, x_n,  \frac{1}{n} \sum x_i \right| y\right)^\prime
\end{equation}

we find $A_{n+1} v_{n+1} = \lambda v_{n+1}$.

Now suppose $u_{n+1} = (x_1,\ldots,x_{n+1} | y)'$ and suppose $A_{n+1} u_{n+1} = \lambda u_{n+1}$. Taking

\begin{equation}
u_n = \left( \left. \frac{n x_1 + x_{n+1}}{n+1}, \ldots, \frac{n x_n + x_{n+1}}{n+1}  \right| y \right)^\prime
\end{equation}

we find $A_n u_n = \lambda u_n$. 

Finally, we must consider the possibility that $u_n = 0$, i.e.~$x_i = - x_{n+1}/n$ $\forall i$. If $n \geq 2$, this may be overcome by first creating a new eigenvector $u_{n+1}^\prime = (x_{n+1},x_2,\ldots,x_n,x_1 | y)$ by swapping two elements, then constructing $u_n$ as above. Thus we may conclude $\lambda(A_n) \supseteq \lambda(A_{n+1})$ for $n \geq 2$ as required.

The practical implication of the above theorem is that, the dynamic behaviour (or at least the linear dynamic behaviour) of a full system of cells may be captured by bulking the cells into two compartments. If cells are instead bulked as one, some behaviour will be lost. 

Moving back to specifics, we may construct the two sets of eigenvectors  associated with the system. If $u_1 = (x | y)^\prime$ is an eigenvector of $A_1$, then $u_n = (x, \ldots, x | y)^\prime$ is the corresponding eigenvector of $A_n$. If $v = x$ is an eigenvector of $f_x$, then $v_n = (x, 0, \ldots, 0, -x,0, \ldots, 0 | 0)$ are the corresponding eigenvectors of $A_n$.

\section{An example}

From a stability perspective, the system

\begin{equation}
x^\prime = f(x,y^\ast)
\end{equation}

may be naturally unstable at $x = x^\ast$, but this instability may be masked in the model through tight control in $y$ -- leading to the eigenvalues of $A_1$ all having negative real part. However, if the cells are not bulked as one, but rather as two (or more) compartments, the feedback exposes the realities of the system as $A_n$ now inherits positive real part eigenvalues from $f_x$. 

For example, the Brusselator is a model proposed in 1968 for an autocatalytic, oscillating chemical reaction~\cite{prigogin68}. In dimensionless form, dynamics may be written as

\begin{eqnarray}
u^\prime &=& 1 - (b+1) u + a u^2 v \label{eq7}\\
v^\prime &=& b u - a u^2 v\label{eq8}
\end{eqnarray}

\begin{figure}
\centering
\resizebox{\figWidth}{!}{%
\includegraphics{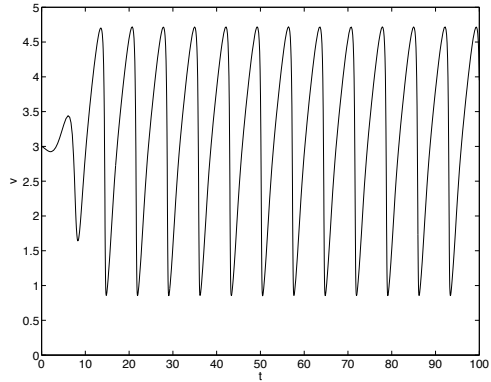}}
\caption{(From Eqs.~(\ref{eq7})--(\ref{eq8})). Stable limit cycle of the Brusselator. Parameter values used are $a=1$, $b=3$, $u(0)=1.01$ and $v(0)=b/a$.}
  \label{fig1}
\end{figure}

Its steady-state is given by $(u,v) = (1,b/a)$ and if $b > a + 1$ there exists a globally-stable limit-cycle (see Fig.~\ref{fig1}).

This model may be transformed by setting $x = (u,v)$ and letting $y=b$ now be a variable representing the externally-supplied nutrient (similar results may be obtained by setting $y=a$).

\begin{eqnarray}
u_i^\prime &=& 1 - (b+1) u_i + a u_i^2 v_i \label{eq10}\\
v_i^\prime &=& b u_i - a u_i^2 v_i \label{eq11}\\
b^\prime &=& g - \frac{1}{n}\sum_i \left(h_1 u_i + h_2 v_i + h_3 b \right)\label{eq14}
\end{eqnarray}

\begin{figure}
\centering
(a) \resizebox{\figWidth}{!}{\includegraphics{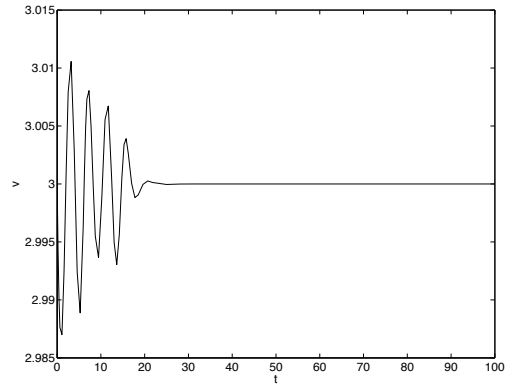}} \\
(b) \resizebox{\figWidth}{!}{\includegraphics{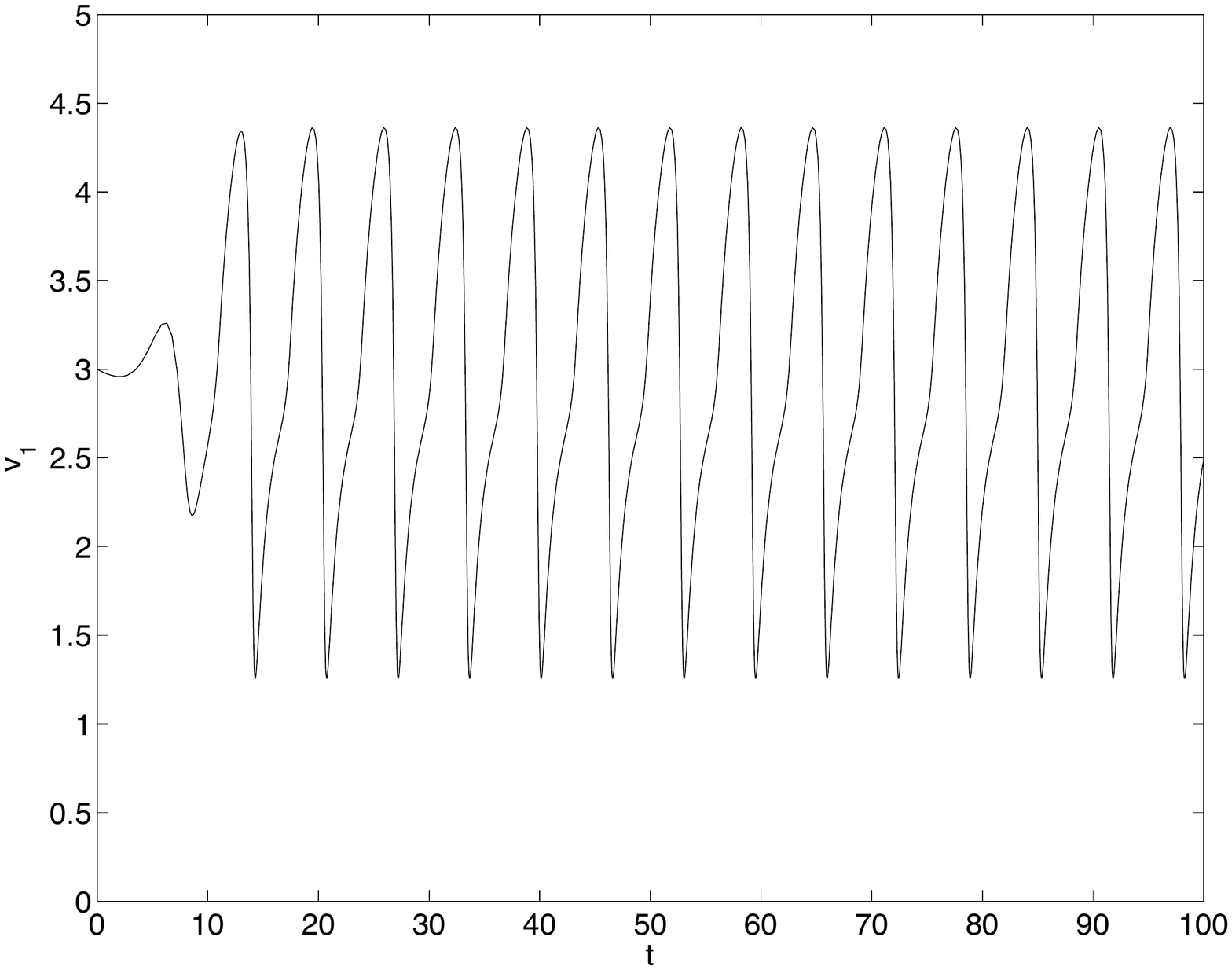}} \\
(c) \resizebox{\figWidth}{!}{\includegraphics{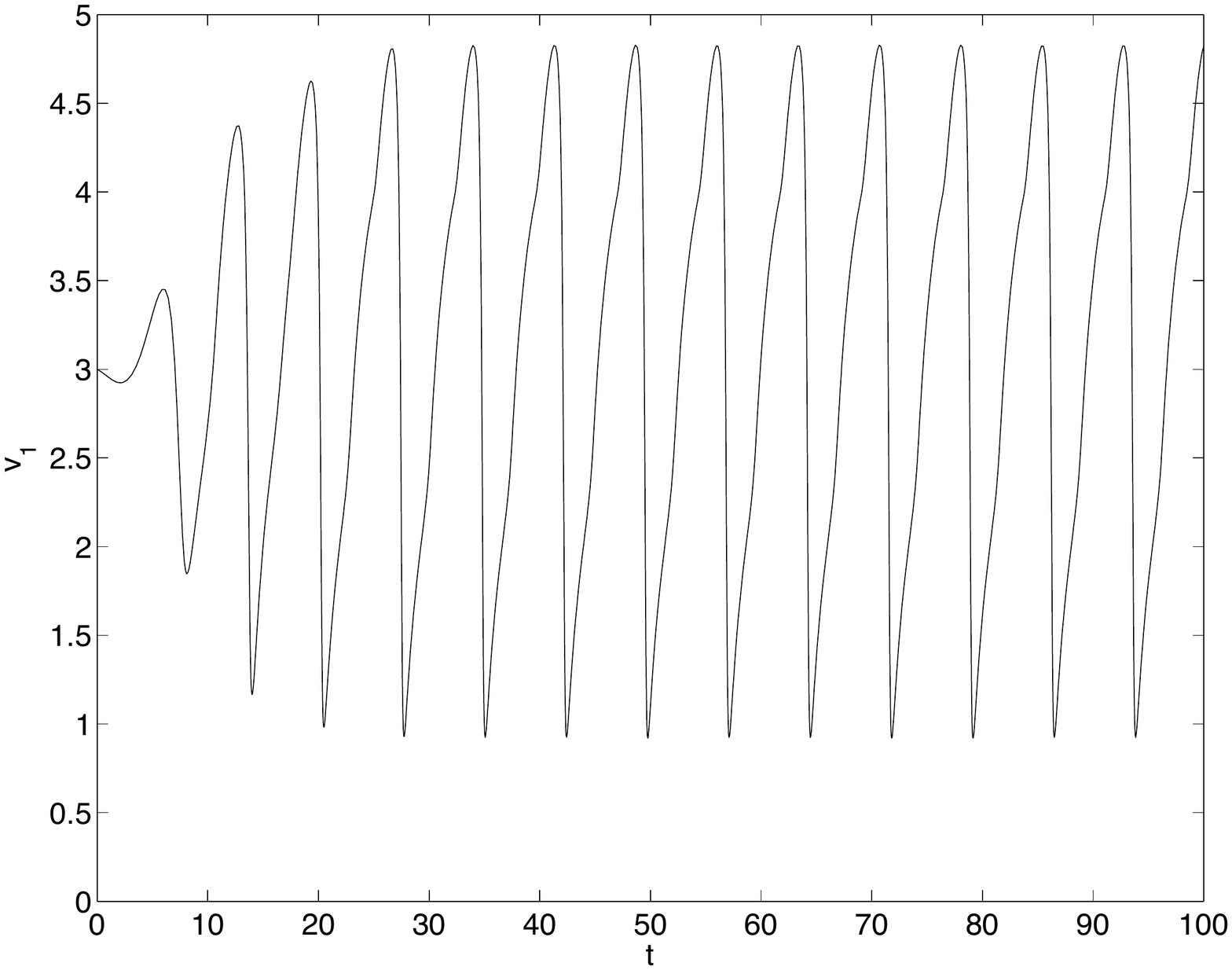}}
\caption{(From Eqs.~(\ref{eq10})--(\ref{eq14})). (a) $n=1$: steady state stabilisation. Parameter values used are as in Fig.~\ref{fig1}, with $g= 2$, $h_1 = -4 $, $h_2 = 0 $, $h_3 = 2$ and $b(0) = 3$. (b) $n=2$: stable limit cycle obtained by dividing populations. Parameter values used are as before, with initial conditions $u_1(0)=1.01$, $u_2(0) = 0.99$. (c) $n=3$: initial conditions $u_1(0)=1.01$, $u_2(0) =1$ and $u_3(0) = 0.99$.}
  \label{fig2}
\end{figure}

For certain parameter values, control on $b$ will seem to stabilise the system (n=1). (see Fig.~\ref{fig2}~(a)). However, when the bulked cells are split,  the underlying oscillations return~(b). Similar dynamics are observed when comparing $n=2$ and $n=3$~(c).

\section{Discussion}

Returning to Fig.~\ref{fig0}, we see the two scales of granularity typically used in metabolic modelling. Typically one represents a population of cells as a single compartment, rather than considering the dynamics of n individual cells. The reasons for this are not clear. It may be that it is assumed that a population of clonal cells would behave in the same way as this. Alternatively, it may be assumed that in order to capture the interactive dynamics, around n times as many differential equations would be required.

As we have shown, both mathematically and via the example of the Brusselator, neither of these assumptions are true. Rather, to answer the titular question, two lumps are required. It is hoped that by using this methodology as standard, new dynamics may be exposed that were previously hidden by the standard assumptions.

\paragraph{Acknowledgements}

I acknowledge the support of the BBSRC/EPSRC Grant BB/ C008219/1 ``The Manchester Centre for Integrative Systems Biology (MCISB)''. Thanks to Dave Broomhead for fruitful discussions.

\end{document}